# Silicon photonics integrated dynamic polarization controller for continuous-variable quantum key distribution

XUYANG WANG,[1,2,7] YANXIANG JIA,[1] XUBO GUO[1], JIANQIANG LIU[1], SHAOFENG WANG[3], WENYUAN LIU[4], FANGYUAN SUN[5], JUN ZOU[6], AND YONGMIN LI [1,2,*]

[1]*State Key Laboratory of Quantum Optics and Quantum Optics Devices, Institute of Opto-Electronics, Shanxi University, Taiyuan 030006, P. R. China*
[2]*Collaborative Innovation Center of Extreme Optics, Shanxi University, Taiyuan 030006, P. R. China*
[3]*College of Physics and Electronic Engineering, Shanxi University, Taiyuan 030006, P. R. China*
[4]*School of Science, North University of China, Taiyuan 030051, P. R. China*
[5]*School of Electrical and Electronic Engineering, Nanyang Technological University, Singapore 639798, Singapore*
[6]*College of Science, Zhejiang University of Technology, Hangzhou 310023, P. R. China*
[7]*wangxuyang@sxu.edu.cn*
*\*yongmin@sxu.edu.cn*

**Abstract:** We designed and demonstrated experimentally a silicon photonics integrated dynamic polarization controller which is a crucial component of a continuous-variable quantum key distribution system. By using a variable step simulated annealing approach, we achieve a dynamic polarization extinction ratio greater than 25 dB. The dynamic polarization controller can be utilized in silicon photonics integrated continuous-variable quantum key distribution system to minimize the size and decrease the cost further.

## 1. Introduction

Quantum communication based on the quantum key distribution (QKD) is well known for its exceptional security, which relies on the quantum uncertainty principle and quantum no-cloning theorem. It is expected to have a wide range of applications in national defense, commerce, and other areas. There are a number of protocols for QKD, which are predominantly divided into discrete variable (DV) QKD and continuous-variable (CV) QKD. Both protocols have their sets of benefits [1–5]. In DV QKD domain, a longer key distribution distance can be achieved [5] and several experiments based on integrated chips have been reported [6-10]. The CV-QKD protocol has satisfactory compatibility with classical optical communication technology [11–19]. It is also compatible with the silicon photonic technology since its optical components are largely classic telecom optical components [20]. In 2019, a CV-QKD based on silicon photonics integrated chips over a 2-m fiber link was reported [21]. This work paves the way for CV-QKD based on integrated silicon photonics chips. However, the key CV-QKD component, dynamic polarization controller (DPC), was not included in their chips. Instead, a manual polarization controller was inserted to modify the state of polarization (SOP). The SOP of the signal and local oscillator (LO) beams could be rotated due to the birefringence of long-distance fibers. At the receiver, the polarization controller should return the SOP to linear polarization. Then the polarization beam splitter (PBS) is employed to separate the signal and LO, and each one enters a PM fiber for further homodyne or heterodyne detection. Note that the QKD system is unattended, therefore the disturbed SOP should be recovered automatically. To this end, the DPC is an essential component.

Several promising platforms on integrated-optic polarization controllers have been reported [22–25]. The architectures of first two works are not based on silicon-on-insulator (SOI) platform, which is compatible with the complementary metal-oxide-semiconductor fabrication technology, more compact and economical. The design in Ref. [24] have many functions. However, the structure was complex, and the insertion loss is large. The design in Ref. [25] cannot realize the endless polarization control. Especially, the DPC based on simulated annealing method was not explored, and the results of dynamic extinction ratio were not obtained [19,20]. In a word, both are not specially designed for CV QKD system. Notice that the polarization controller structures on integrated chips were used to prepare four polarization states required for polarization encoded BB84 protocol, but the DPC has not been demostrated. In this work, we design and demonstrate a silicon photonics integrated DPC. By employing the variable step simulated annealing approach, we achieves a dynamic polarization extinction ratio greater than 25 dB. This device is ready to be integrated into the CV-QKD system that runs on the chip.

In Sec. II, the design of silicon photonics integrated DPC is presented. The overall size of DPC on chip was 2.830 mm × 0.210 mm × 1 mm, which was much smaller than that of fiber-squeezers based DPC. In Sec. III, the characters of the thermal phase shifters (TPS) are evaluated. There is a good linearity between the phase shift and the power consumed by the metal heater. The modulation bandwidth is BW = 30 kHz × 5.6V = 168 kHz·V which is 100 times higher than the modulation speed of the state of the art fiber-squeezers based DPC Porarite III. In Sec. IV, we presents the numerical simulation and experimental results of the silicon photonics integrated DPC. The advantages of the variable step simulated annealing approach is demonstrated anda dynamic polarization extinction ratio larger than 25 dB are obtained. At last, Sec. V presents the conclusion of this paper.

## 2. The design of silicon photonics integrated DPC

Generally, the DPC comprising three or four fiber squeezers were utilized in the CV-QKD system. The fiber squeezers were activated by piezoceramic actuators, which were driven by the high voltage of above 100 V. The squeezers alternated between 0°, 45°, 0°, and 45° of orientation. Each squeezer introduced a phase shift between the linear polarization components aligned parallel and perpendicular to the squeezing direction. The phase shift can be varied by altering the squeezing force. The fiber under the stress can be represented in Jones calculus by two kinds of transformation matrices $M_0$ and $M_{45}$, as indicated in Eq. (1)

$$M_0 = \begin{pmatrix} e^{-i\delta_0/2} & 0 \\ 0 & e^{i\delta_0/2} \end{pmatrix}, M_{45} = \begin{pmatrix} \cos(\delta_{45}/2) & -i\sin(\delta_{45}/2) \\ -i\sin(\delta_{45}/2) & \cos(\delta_{45}/2) \end{pmatrix}, \quad (1)$$

where $\delta_0$ and $\delta_{45}$ are the delayed phases due to stressed fiber birefringence. The matrix $M_{45}$ can be transformed to a product of matrix $M_0$ and matrices which describe 50/50 couplers as follows.

$$M_{45} = \begin{pmatrix} \cos(\delta_{45}/2) & -i\sin(\delta_{45}/2) \\ -i\sin(\delta_{45}/2) & \cos(\delta_{45}/2) \end{pmatrix} = \begin{pmatrix} 1/\sqrt{2} & -1/\sqrt{2} \\ 1/\sqrt{2} & 1/\sqrt{2} \end{pmatrix} \begin{pmatrix} e^{-i\delta_{45}/2} & 0 \\ 0 & e^{i\delta_{45}/2} \end{pmatrix} \begin{pmatrix} 1/\sqrt{2} & 1/\sqrt{2} \\ -1/\sqrt{2} & 1/\sqrt{2} \end{pmatrix} \quad (2)$$

The transformation matrices can be transformed into the corresponding structures of silicon photonics integrated circuits, as demonstrated in Figure 1 [26]. The blue lines denote the waveguides. The red and black rounded rectangles denote the TPS and 50/50 multimode interferences (MMI), respectively.

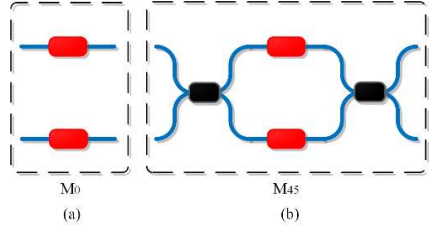

Fig.1 The structures of silicon photonics integrated circuits corresponding to the transformation matrices $M_0$ and $M_{45}$. (a) The structure corresponds to the matrix $M_0$; (b) The structure corresponds to the matrix $M_{45}$.

According to the above basic structures, the DPC consisting of silicon photonics integrated circuits were designed, as indicated in Figure 2 (a). The input and output ports were 2D grating couplers (GC) that were utilized as the PBS [27]. The structure was symmetrical, the input port can be utilized as an output port, and vice versa. Four 1×2 50/50 MMI couplers were inserted into the waveguides for the purpose of testing and aligning, where one output port of each 1×2 MMI coupler is connected to an 1D GC or a photodiode. The 1550 nm laser beam in single-mode fiber with an arbitrary SOP can be guided into the integrated DPC by a vertical coupling approach. Each linear polarization component is directed into one waveguide with TE mode using the 2D GC. Following a series of transformations, such as $M1(M_0)$, $M2(M_{45})$, $M3(M_0)$, and $M4(M_{45})$, the two TE modes of each waveguide are coupled out to another singlemode fiber through the 2D GC.

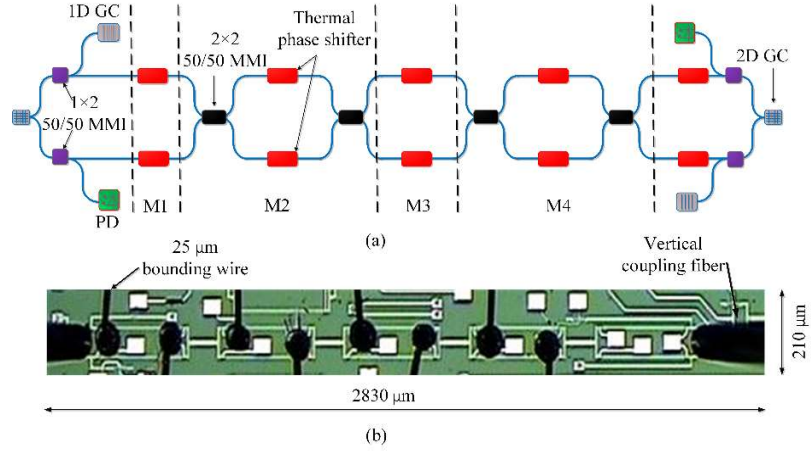

Fig.2 The structure and photograph of silicon photonics integrated DPC. PD, photodiode; (a) The structure of DPC; (b) The photograph of DPC.

Figure 2 (b) indicates the photograph of silicon photonics integrated DPC. The fabrication of the device was performed with CSiP180Al active flow technology. It was based on a 200 mm SOI substrate with 2 μm BOX and 220 nm top silicon. The overall size of DPC was 2.830 mm × 0.210 mm × 1 mm, which was much smaller than that of fiber-squeezers based DPC with the size of 83 mm × 20.32 mm × 16 mm (polarite III, GP). Here, TPS was employed to alter the delayed phase. The 25 μm golden wire was selected to bond the pad by which the driving voltage can be applied on the DPC. The characters of the TPS are evaluated in detail in the next section.

### 3. The characterization of the TPS

The characters of the TPS or TiN metal heater are evaluated utilizing an MZ modulator. The resistance of the metal heater is 1.97k and 1 mW 1550 nm continuous laser beam was coupled into one port of the 2×2 MZ modulator. By varying the voltage on the metal heater of one side, the output power was altered. We transformed the voltage $V$ into the power $P$ applied on the metal heater utilizing the expression

$$P = V^2 / R, \qquad (3)$$

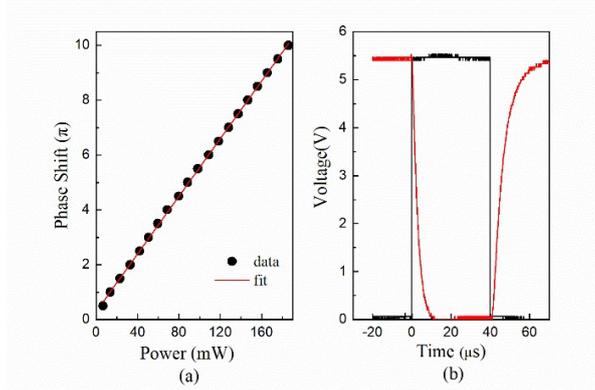

Fig. 3 The characterization of the TPS. (a) The phase shift versus the consumed power of the thermal heater; (b) The rise and fall time of the MZ modulator.

where $R$ denotes the resistance of the metal heater.

As shown in Figure 3(a), there is a good linearity between the phase shift and the power consumed by the metal heater. The total phase shift is roughly $3\pi$ by varying the the power from 0 to 50.56 mW, which corresponds to the applied voltage from 0 to 10 V. The phase shift amount isenough to stabilize the polarization with the simulated annealing method. Note that many digital-to-analog converterchip's output voltages can reach this voltage and a high voltage amplifier is unnecessary. In contrast, the voltage of 140 V is required to drive the piezo of Polarite III.

In addition, the modulation speed of the phase was evaluated, which determines the polarization locking speed. Furthermore, it determined the ability of the CV QKD system to resist the polarization disturbance of quantum channel [28]. In the measurement, a 40 μs electronic square pulse with a peak voltage of 5.6 V was applied to the metal heater of the MZ modulator. It correspondes to a phase shift of $\pi$. The rising time was measured to be 11 μs and the falling time is 5.9 μs. This value is about one-tenth of the rise time of Polarite III. Thus the modulation bandwidth is about 30 kHz.

## 4. Numerical simulation and experimental results

### 4.1 The method of varied step simulated annealing

The ability to perform high-speed polarization control and obtain a high polarization extinction ratio is very crucial in CV QKD [28]. Simulated annealing approach is a good approach to lock the output SOP. Usually, a big step is needed to realize high-speed polarization controlling. However, from the simulation, we discovered that to obtain a higher polarization extinction ratio, a small step is necessary. To address this issue, the variable step approach was adopted to increase the speed and extinction ratio of the SOP locking simultaneously. The flow chart of the program is illustrated in Figure 4.

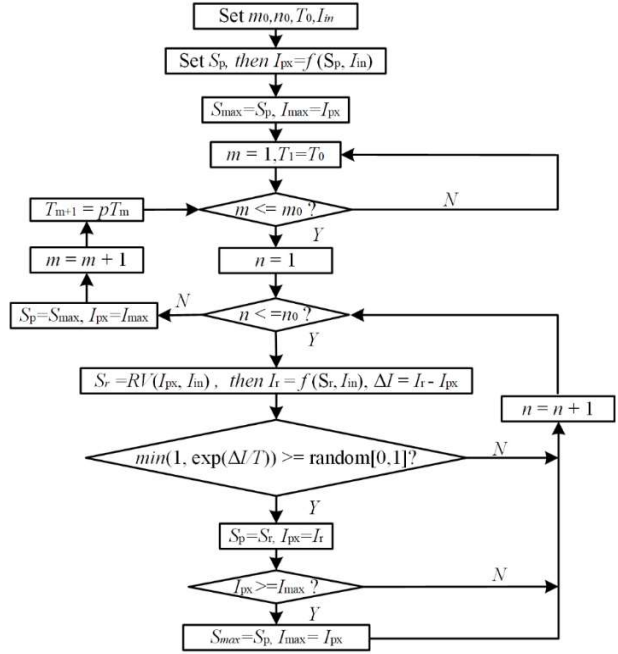

Fig.4 The flow chart of the varied step simulated annealing method.

Due to the sensitivity of the quantum state to the loss in the signal path, a 90/10 polarization maintaining coupler was usually inserted in the LO path to control the SOP [13]. The simulated annealing approach was utilized to maximize the intensities of the beam in the LO path. We started by setting the initial temperature $T_0 = 10^{-5}$, the number of external circulation loops $m_0 = 10$, and internal circulation loops $n_0 = 50$, and the intensity of the input beam $I_{in}$ to a normalized value of 1. Then we set the initial value of the four delay phases to half of the maximum of total phase shift and stored them into an array $S_p(\theta_1, \theta_2, \theta_3, \theta_4)$. Then we calculated the output intensity $I_{px}$ according to the delay phases and the SOP of the input beam. Here, $I_{px}$ denotes the intensity of LO from one output port of the PBS follows the DPC. $I_{max}$ and $S_{max}$ denotes the maximum intensity of $I_{px}$ and corresponding delayed phases, respectively. $I_{py}$ denotes the intensity of signal path from the other output port.

Before entering the external circulation loops, the value $m$ was initialized to 1 and the value $T_1$ was initialized to $T_0$. In the external circulation loop, the temperature decreases with the proportion function

$$T_{n+1} = pT_n. \tag{4}$$

When implementing the internal circulation loop, we employed the random variable step function $RV(I_{px}, I_{in})$ to generate random phase delay $S_r$. The random variable step function is expressed as follows:

$$S_r = \begin{cases} S_p + st \cdot r, & S_p \leq 0 \\ S_p + c \cdot st \cdot r, & 0 < S_p < 3\pi \\ S_p - st \cdot r, & 3\pi \leq S_p \end{cases}, \tag{5}$$

where $r$ denotes the random numbers between 0~1, $c$ denotes the random numbers of −1 or 1, and $st$ represents the search step. Usually, when the step $st$ is larger, the search time is shorter. However, the disturbance generated by the large step $st$ is also higher, making a high extinction ratio of SOP difficult to obtain. We proposed to use a variable step $st$ to improve the extinction ratio, and in the mean time maintain a high search speed. The variable steps are indicated by

$$st = \begin{cases} 0.16 rad, & 0.1 < I_{st} <= 1 \\ 0.08 rad, & 0.01 < I_{st} <= 0.1 \\ 0.03 rad, & 0.001 < I_{st} <= 0.01 \\ 0.008 rad, & I_{st} <= 0.001 \end{cases}. \tag{6}$$

where $I_{st} = 1 - I_{px}$ defines the gap between the normalized maximum intensity 1 and $I_{px}$. The simulation results are shown in Figure 5. The fixed step method with two different steps and the variable step method were compared. When the step was larger (st = 0.16 rad), the required internal circulation times was less. However, the extinction ratio is lower. For a smaller step of $st = 0.008$, a higher extinction ratio can be obtained at the cost of a longer time to reach the steady value. It is clear that the best results is achieved by the varied step simulated annealing methods, where the highest extinction ratio is obtained at an least internal circulation times.

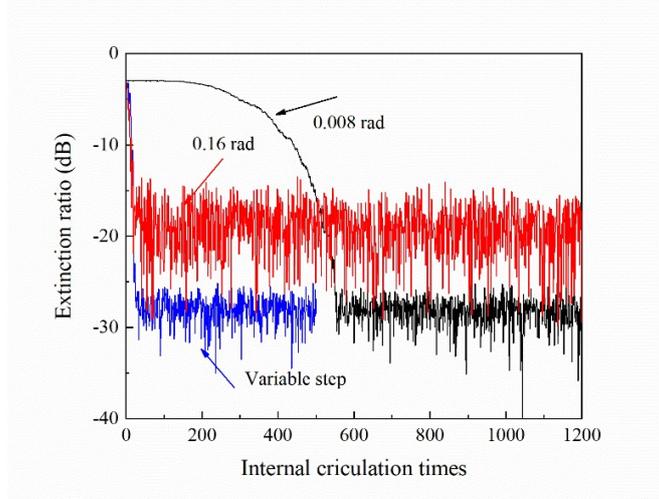

Fig.5 The simulation results of dynamic polarization controlling. It represents the extinction ratio versus the internal circulation times with fixed and varied step simulated annealing methods.

Note that the static polarization extinction ratio, electronic noises of the photodectors, and the fluctuations of the optical power determine the ultimate limit of the DPC. In the simulation, the minimum intensiy of $I_{py}$ is set to $I_{px} \cdot 10^{-2.8}$ considering the statistic polarizaiotn extionction ratio of the 2D GC of 28 dB. The standard variation of the electronic noises and optical power fluctuation is set to $5 \times 10^{-4}$.

### 4.2 Experimental results

The scheme of the test setup is illustrated in Figure 6. A 1550 nm fiber pigtailed DFB laser was utilized to generate a continuous beam, and a variable optical attenuator is used to tune the intensity of the laser beam. Part of the beam, which was employed to monitor the laser power, was separated by a 50/50 fiber single-mode (SM) coupler. A manual polarization controller is employed to tune the SOP of the beam directed into the chip. An alignment machine was utilized to align the fiber to the top of the 2D GC nearly vertically with an angle of 80°. The power coupled into the chip is 1 mW. The measured total loss from the input 2D GC to the output 2D GC is ~20 dB. The coupling loss of the fabricated 2D GC is 7 dB. Considering the 3dB loss from the 50/50 MMI on the chip, the transmission loss of the integrated DPC is 3 dB which mainly originates from the four MMIs. After transmitting through the chip, the beam was directed into the SM fiber again. Then, it was separated by the fiber PBS into path 1 and path 2. The intensity of the beam in the paths was evaluated byDetector 1 and Detector 2. The DPC on the chip is exploited to maximize the intensity of the beam in path 1 (Detector 1) and therefore the intensity of the beam in path 2 (Detector 2) is minimized.

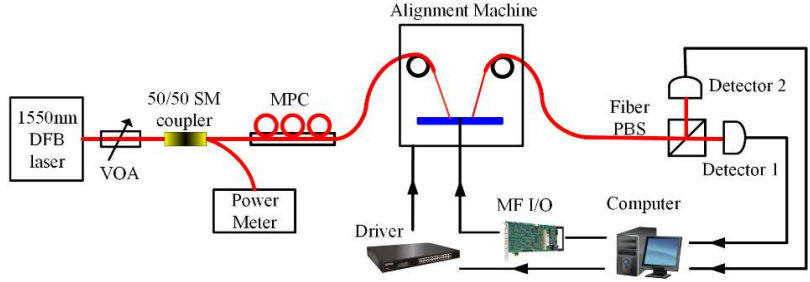

Fig.6. Experimental setup to test the silicon photonics integrated DPC. MPC, Manual polarization controller; MF, Multifunctional.

Based on the simulation results in Sec. 4.1, we employed the the variable step approach to lock the SOP of the laser. The output voltage limit of multifunction I/O card was roughly 10 V. From Sec. 3, this voltage range can results in a phase shift of 3π. According to the simulation results, this phase shift range is enough to lock the SOP by setting the initial phase shift at 3π. By fitting the experiment results in Figure 3 (a), a linear formula can be obtained as follows

$$\theta = c \cdot P + \theta_{bias},  \quad (7)$$

where $c = 164.85$ rad/mW, $\theta_{bias} = 0.93$ rad. Combined with Eq. (3), the relationship between the voltage step $\Delta V$ and corresponding phase step $\Delta \theta$ can be written as

$$\Delta V = \frac{R}{2cV} \Delta \theta. \quad (8)$$

From Eq. (8), we observe that $\Delta V$ decreases with the voltage $V$ for a constant $\Delta \theta$. When the value of voltage $V$ is small, the voltage step $\Delta V$ will be very big. The minimum value $\Delta V_{min}$ corresponds to the maximum value of $V_{max}$, and it is not relevant to $\theta_{bias}$. In our experiment, $\Delta V_{min}$ was used to ensure a stabler polarization controlling for corresponding $\Delta \theta$ or $st$ value.

The experiment results of our silicon photonics integrated dynamic polarization controlling are shown in Fig. 7. Before the operation of DPC control procedure, the voltage of detector 2 was set to the maximum value by MPC. For the measurement range of detector 2, its output was saturated at the beginning of DPC control. The black line represents the results utilizing a fixed step of 0.005 V with ~600 internal circulation times. The red line represents the results utilizing a fixed step of 0.1V with ~100 internal circulation times. The blue line represents the results utilizing a varied step. For the varied step approach, a high extinction ratio higher than 25 dB is obtained with ~100 internal circulation times. This value is nearly close to the static extinction ratio of 28 dB for our silicon photonics integrated DPC. The experimental results is approximately in agreement with the simulation results in Fig. 5.

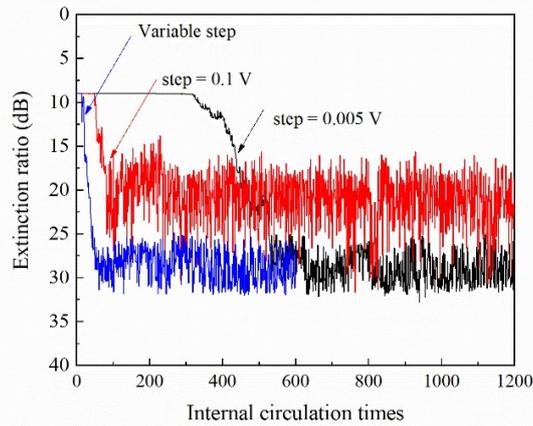

Fig.7 The experiment results of dynamic polarization controlling. It represents the extinction ratio versus the internal circulation times with fixed and varied step simulated annealing methods.

## 5. Conclusions

We designed and demonstrated a silicon photonics integrated dynamical polarizatioin controller. A polarization extinction ratio of 25 dB were achieved with the variable step simulated annealing approach. This device is suitable for silicon photonics integrated CV QKD systems. It is expected the device can also be used for reference in silicon photonics integrated DV QKD systems. Our current silicon photonics integrated DPC have roughly 10 dB insertion loss including 7 dB coupling loss plus 3 dB of transmission loss on chip. Note that any insertion loss in the quantum channel will be detrimental to the QKD system. For further study, it is very promising to decrease the coupling efficiency from 7 dB to 2 dB or less by using butting coupling method with an appropriate spot size converter between the on-chip silicon waveguide and single mode fiber [29]. In this case, the effective transmission loss will be significantly reduced resulting in a better QKD system performance.

**Funding.** This work has been supported by the Aeronautical Science Foundation of China (20200020115001); Key Research and Development Program of Guangdong Province (2020B0303040002); Shanxi 1331KSC; National Natural Science Foundation of China (11774209); the Open Project of the State Key Laboratory of Quantum Optics and Quantum Optics Devices of Shanxi University under Grant KF202006.

**Disclosures.** The authors declare that there are no conflicts of interest related to this article.

**Data availability.** The data that support the findings of this study are available from the corresponding author upon reasonable request.

## 6. References

1. V. Scarani, H. Bechmann-Pasquinucci, N. J. Cerf, M. Dušek, N. Lütkenhaus, and M. Peev, "The security of practical quantum key distribution," Rev. Mod. Phys. **81**(3), 1301–1350 (2009).
2. C. Weedbrook, S. Pirandola, R. García-Patrón, N. J. Cerf, T. C. Ralph, J. H. Shapiro, and S. Lloyd, "Gaussian quantum information," Rev. Mod. Phys. **84**(2), 621–669 (2012).

3. S. Pirandola, R. Laurenza, C. Ottaviani, and L. Banchi, "Fundamental limits of repeaterless quantum communications," Nat. Commun. **8**, 15043 (2017).
4. M. Lucamarini, Z. L. Yuan, J. F. Dynes, and A. J. Shields, "Overcoming the rate–distance limit of quantum key distribution without quantum repeaters," Nature **557**, 400–407 (2018).
5. M. Minder, M. Pittaluga, G. L. Roberts, M. Lucamarini, J. F. Dynes, Z. L. Yuan, and A. J. Shields, "Experimental quantum key distribution beyond the repeaterless secret key capacity," Nat. Photonics 13, 839 (2019).
6. K. Wei, W. Li, H. Tan, Y. Li, H. Min, W. J. Zhang, H. Li, L. X. You, Z. Wang, X. Jiang, T. Y. Chen, S. K. Liao, C. Z. Peng, F. H. Xu, and J. W. Pan, "High-speed measurement-device-independent quantum key distribution with integrated silicon photonics," Phys. Rev. X **10**(3), 031030 (2020).
7. C. X. Ma, W. D. Sacher, Z. Y. Tang, J. C. Mikkelsen, Y. Yang, F. H. Xu, T. Thiessen, H. K. Lo, J. K. S. Poon, "Silicon photonic transmitter for polarization-encoded quantum key distribution," Optica **3**(11), 1274-1278 (2020).
8. P. Sibson, J. E. Kennard, S. Stanisic, C. Erven, J. L. O'Brien, and M. G. Thompson, "Integrated silicon photonics for high-speed quantum key distribution," Optica **4**(2), 172-177 (2017).
9. P. Sibson, C. Erven, M. Godfrey, S. Miki, T. Yamashita, M. Fujiwara, M. Sasaki, H. Terai, M.G. Tanner, C. M. Natarajan, R.H. Hadfield, J.L. O'Brien & M.G. Thompson, "Chip-based quantum key distribution," Nat. Commun. **8**, 13984 (2017).
10. J. W. Wang, F. Sciarrino, A. Laing, and M. G. Thompson, "Integrated photonic quantum technologies," Nat. Photonics **14**, 273-284 (2020).
11. F. Grosshans, G. Van Assche, J. Wenger, R. Brouri, N. J. Cerf, and P. Grangier, "Quantum key distribution using Gaussian-modulated coherent states," Nature **421**, 238–241 (2003).
12. J. Lodewyck, M. Bloch, R. García-Patrón, S. Fossier, E. Karpov, E. Diamanti, T. Debuisschert, N. J. Cerf, R. Tualle-Brouri, S. W. McLaughlin, and P. Grangier, "Quantum key distribution over 25 km with an all-fiber continuous-variable system," Phys. Rev. A **76**(4), 042305 (2007).
13. P. Jouguet, S. Kunz-Jacques, A. Leverrier, P. Grangier, and E. Diamanti, "Experimental demonstration of long-distance continuous-variable quantum key distribution," Nat. Photonics 7, 378–381 (2013).
14. R. Kumar, H. Qin, and R. Alléaume, "Coexistence of continuous-variable QKD with intense DWDM classical channels," New J. Phys. **17**(4), 043027 (2015).
15. C. Wang, D. Huang, P. Huang, D. Lin, J. Peng, and G. Zeng, "25 MHz clock continuous-variable quantum key distribution system over 50 km fiber channel," Sci. Rep. **5**, 14607 (2015).
16. B. Qi, P. Lougovski, R. Pooser, W. Grice, and M. Bobrek, "Generating the local oscillator "Locally" in continuous-variable quantum key distribution based on coherent detection," Phys. Rev. X **5**(4), 041009 (2015).
17. D. Huang, P. Huang, H. Li, T. Wang, Y. Zhou, and G. Zeng, "Field demonstration of a continuous-variable quantum key distribution network," Opt. Lett. **41**(15), 3511–3514 (2016).
18. X. Y. Wang, W. Y. Liu, P. Wang, and Y. M. Li, "Experimental study on all-fiber-based unidimensional continuous-variable quantum key distribution," Phys. Rev. A **95**(6), 062330 (2017).
19. Y. C. Zhang, Z. Y. Chen, S. Pirandola, X. Y. Wang, C. Zhou, B. J. Chu, Y. J. Zhao, B. J. Xu, S. Yu, and H. Guo, "Long-distance continuous-variable quantum key distribution over 202.81 km of fiber," Phys. Rev. Lett. **125**, 010502 (2020).
20. E. Diamanti, H. K. Lo, B. Qi, and Z. Yuan, "Practical challenges in quantum key distribution," npj Quant. Inf. **2**, 16025 (2016).
21. G. Zhang, J. Y. Haw, H. Cai, F. Xu, S. M. Assad, J. F. Fitzsimons, X. Zhou, Y. Zhang, S. Yu, J. Wu, W. Ser, L. C. Kwek, and A. Q. Liu, "An integrated silicon photonic chip platform for continuous-variable quantum key distribution," Nat. Photonics **13**, 839-842 (2019).
22. J. D. Sarmiento-Merenguel, R. Halir, X. Le Roux, C. Alonso-Ramos, L. Vivien, P. Cheben, E. Durán-Valdeiglesias, I. Molina-Fernández, D. Marris-Morini, D. X. Xu, and J. H. Schmid. "Demonstration of integrated polarization control with a 40 dB range in extinction ratio," Optica **2**(12), 1019–1023 (2015).
23. J. W. Kim, S. H. Park, W. S. Chu, and M. C. Oh, "Integrated-optic polarization controllers incorporating polymer waveguide birefringence modulators," Opt. Express, **20**(11), 12443–12448 (2012).
24. H. L. Zhou, Y. H. Zhao, Y. X. Wei, F. Li, J. J. Dong, and X. L. Zhang, "All-in-one silicon photonic polarization processor," Nanophotonics **8**(12), 2257–2267 (2019).
25. P. Velha, V. Sorianello, M. V. Preite, G. De Angelis, T. Cassese, A. Bianchi, F. Testa, and M. Romagnoli. "Wide-band polarization controller for Si photonic integrated circuits," Opt. Lett, **41**(24), 5656–5659 (2016).
26. L. Moller, "WDM Polarization Controller in PLC Technology," IEEE Photon. Technol. Lett. **13**(6), 585–587 (2001).
27. J. Zou, Y. Yu, X. L. Zhang, "Two-dimensional grating coupler with a low polarization dependent loss of 0.25 dB covering the C-band," Opt. Lett. **41**(18), 4206–4209 (2016).
28. W. Y. Liu, Y. X. Cao, X. Y. Wang, and Y. M. Li, "Continuous-variable quantum key distribution under strong channel polarization disturbance," Phys. Rev. A **102**(3), 032625 (2020).
29. J. Zou, X. Ma, X. Xia, C. H. Wang, M. Zhang, J. H. Hu, X. Y. Wang, and Jian-Jun He, "Novel wavelength multiplexer using (N + 1) × (N + 1) arrayed waveguide grating and polarization-combiner-rotator on SOI platform," J. Lightwave Technol. **39**(8), 2431–2437 (2021).